\documentstyle[prl,aps,epsf]{revtex}
\draft
\begin{document}
\twocolumn[\hsize\textwidth\columnwidth\hsize\csname @twocolumnfalse\endcsname

\title{Incommensuration Effects and Dynamics in Vortex Chains} 
\author{C. Reichhardt$^{\rm (1)}$ and C. J. Olson Reichhardt$^{\rm (2)}$}
\address{$^{\rm (1)}$CNLS and Applied Physics Division, 
Los Alamos National Laboratory, Los Alamos, NM 87545}
\address{$^{\rm (2)}$ Theoretical and Applied Physics Divisions,
Los Alamos National Laboratory, Los Alamos, NM 87545}

\date{\today}
\maketitle
\begin{abstract}
We examine the motion of one-dimensional (1D) 
vortex matter embedded in a 2D vortex
system with weak pinning using numerical simulations. We confirm the
conjecture of Matsuda {\it et al.} (Science {\bf 294}, 2136 (2001))
that the onset of the temperature induced motion of the chain is due
to an incommensuration effect of the chain with
the periodic potential created by the
bulk vortices. In addition, under an applied driving force we find
a two stage depinning transition, 
where the initial depinning of the vortex chain 
occurs through soliton like pulses. When an 
ac drive is added to the dc drive, we observe
phase locking of the moving vortex chain. 
\end{abstract}
\vspace{-0.1in}
\vspace{-0.3in}

\vskip2pc]
\narrowtext
The static and dynamical behavior of elastic media interacting with
periodic or random substrates is a topic of intense study as it 
relates to a wide variety of physical systems, such as friction where
atoms move over a periodic potential created by the underlying 
stationary atoms\cite{Persson1}, 
sliding charge-density waves \cite{Gruner2}, and vortex motion
in superconductors with random quenched disorder \cite{Giamarchi3} 
or periodic hole or dot arrays
\cite{Moshchalkov4,VanLook5,Surdeanu6,Reichhardt7}. 
A simplified case occurs when the motion 
is confined in 1D or quasi 1D systems. An 
example of this is vortex motion in superconductors 
with nano fabricated arrays of channels \cite{Kes8,Smith9}.  There 
are also naturally occurring defects that can act as quasi-1D channels for 
vortex motion, such as grain \cite{Hog10,Grain11} and 
twin boundaries \cite{Groth12}. The vortices in 
fabricated channels experience two types
of pinning: the first from the intrinsic random disorder in the sample, and
the second from the periodic potential created by the immobile vortex 
lattice surrounding the channels. 
Experiments and simulations 
in these systems have observed 
commensuration effects and interesting dynamical states \cite{Kes8,Smith9}.

Another intriguing example of quasi 1D vortex matter is      
the vortex chain state which occurs in layered superconductors
with tilted magnetic fields as first imaged by Bolle {\it et al.} \cite{Bolle13}.
Here a portion of the vortices align in a string with a smaller
lattice spacing than that of the bulk vortices.
Recently scanning Hall probe \cite{Bending14} and 
Lorentz microscopy \cite{Matsuda15} images have shown that the
vortex chain states exhibit a remarkably rich variety of behaviors and
structures for various regimes. The recent Lorentz microscopy 
experiments found that for low temperatures,
both the vortices along the chain state 
and the bulk vortices are stationary. 
For increasing temperatures
the vortices along the chain disappear, which is speculated to be due to 
the onset of motion or oscillation of the vortex chains, so that the 
chain vortex images are blurred 
while the bulk vortices are still stationary \cite{Matsuda15}. 
It was conjectured in \cite{Matsuda15} that this effect 
is due to an incommensuration
effect where vortices in the chain move in a periodic 
potential created by the bulk vortices. Since the vortices in the chain
have a different spacing than the periodic potential, the chain is 
incommensurate and hence more weakly pinned than if the chain were 
commensurate. 
These images also showed that the blurring first occurs 
where there a mismatch accompanied by a
topological defect in the chain. 
Additional evidence for this scenario is that as the field is increased,
the periodicity of the potential decreases and the incommensuration
effects are enhanced, causing the
disappearance of the chains to shift to lower temperatures,
as observed\cite{Matsuda15}. 
Dodgson \cite{Dodgson16} has proposed an alternative
explanation 
in which the image disappears due the sudden formation of tilted chains. 
In addition to the intriguing physics of the vortex matter in confined
geometries it is also of paramount importance to understand vortices in the
chain state or along grain and twin boundaries, as these can act as
weak lines where flux motion occurs, leading to the breakdown of 
superconductivity in the chains prior to the 
bulk. 

In order to examine the conjecture that the disappearance of the vortex 
chain is due to an incommensuration effect, as well as to explore 
dynamical consequences of these incommensuration effects,
we have performed
numerical simulations of vortex chains coexisting with bulk vortices in the 
presence of weak quenched disorder and thermal noise. We also 
investigate the effects of applied dc and combined dc and ac 
external drives on the chain motion. Our results should be 
relevant for the vortex chain state where the vortices remain
aligned in the $z$-direction so that much of the physics can be considered
2D.  In addition our results are relevant for vortex motion in artificial 
channels and grain or twin boundaries where there are additional vortices
outside the channels.   

We consider a 2D sample with periodic boundary conditions in the
$x$ and $y$ directions. We model the vortices as repulsive point particles
which interact with weak random quenched disorder, applied 
driving forces, and thermal noise. A portion of the vortices are confined to
move along a quasi-1D channel.
The overdamped equation of motion for a vortex $i$ is  
\begin{equation}
\eta\frac{{\bf dr}}{dt} = {\bf f}_{i} = {\bf f}^{vv}_{i} + 
{\bf f}^{rp}_{i} + {\bf f}^{d}_{i} + {\bf f}^{T}_{i} 
\end{equation} 
Here $\eta$ is the damping constant that is set equal to unity.
The vortex-vortex interaction force is 

\begin{figure}
\center{
\epsfxsize=3.5in
\epsfbox{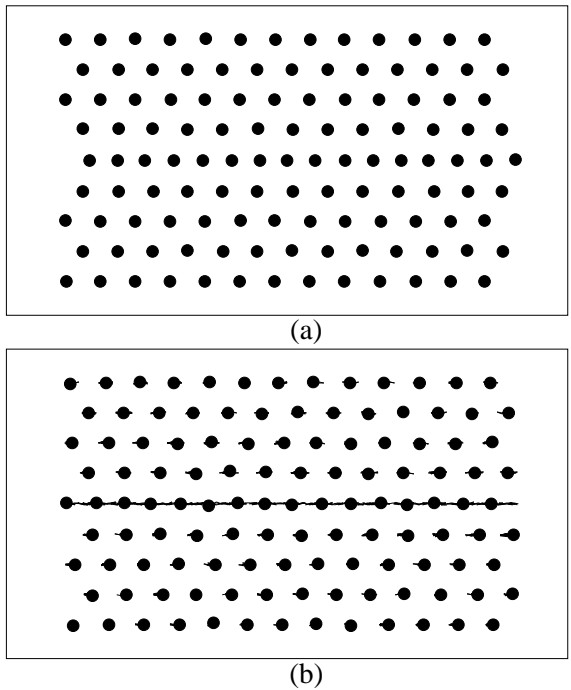}}
\caption{A portion of the sample with  vortices (black dots)
and trajectories (lines). The 
vortex chain is in the center. (a) $T/T_{m} = 0.15$. (b) $T/T_{m} = 0.25$, 
showing that motion is occurring along the chain but not 
in the bulk.   
} 
\label{fig1}
\end{figure}

\hspace{-13pt}
${\bf f}^{vv}_{i} = -\sum_{j\neq i}^{N_{v}}\nabla_i U_{v}(r_{ij})$,
where the vortex interaction potential is $U_{v} = -A_{v}\ln(r)$. 
The force from random quenched disorder ${\bf f}^{rp}_{i}$ comes
from randomly placed attractive parabolic traps of range $r_{p}$ with
$r_{p}/a = 0.1$, where $a$ is the vortex lattice constant and the traps
have a maximum force of $f_{p}=1.0$. The driving forces ${\bf f}^{d}_{i}$ 
include a dc force ${\bf f}^{DC}_{i}$ which is modeled as a uniform
drive on the vortices in the $x$-direction. We start from zero driving 
and with small increments increase the dc driving force. At each increment we
average the vortex velocity $V_{x}$ over a fixed time interval.
An ac drive can also be applied in the
form ${\bf f}^{AC}_{i} = A\sin(\omega t){\bf \hat{x}}$, where $A$ is the 
ac amplitude 
and $\omega$ is the frequency. The
forces from the thermal noise ${\bf f}^{T}_{i}$ come from random Langevin
kicks with $<f^{T}(t)> = 0$ and 
$<f^{T}(t)f^{T}(t^{'})> = 2\eta k_{B}T\delta(t-t^{\prime})$.    
We measure temperature in units of $T_{m}$,  
the temperature at which the bulk vortices melt. For high-$T_{c}$ 
materials at fields where the chain state is observed, 
$T_{m}$ is around $70-80K$
and for the low-$T_{c}$ materials with nano-channels, $T_{m}$ would correspond
to $T_{c}$. 
Lengths are measured in units of $a$,
and forces in $A_{v}$, the vortex-vortex interaction prefactor.    

We first consider the effects of temperature in 
systems with no external driving force to test the conjecture in
\cite{Matsuda15}. 
We start the system at $T/T_{m} = 0.0$ where the bulk vortices
form a triangular lattice with lattice constant $a$. 
Additional
vortices are added along the channel, giving a vortex spacing 
in the channel of $a^{\prime}<a$. 
We monitor the vortex displacements for fixed time intervals of
$10^4$ MD steps, and measure the 
vortex positions and trajectories.

\begin{figure}
\center{
\epsfxsize=3.5in
\epsfbox{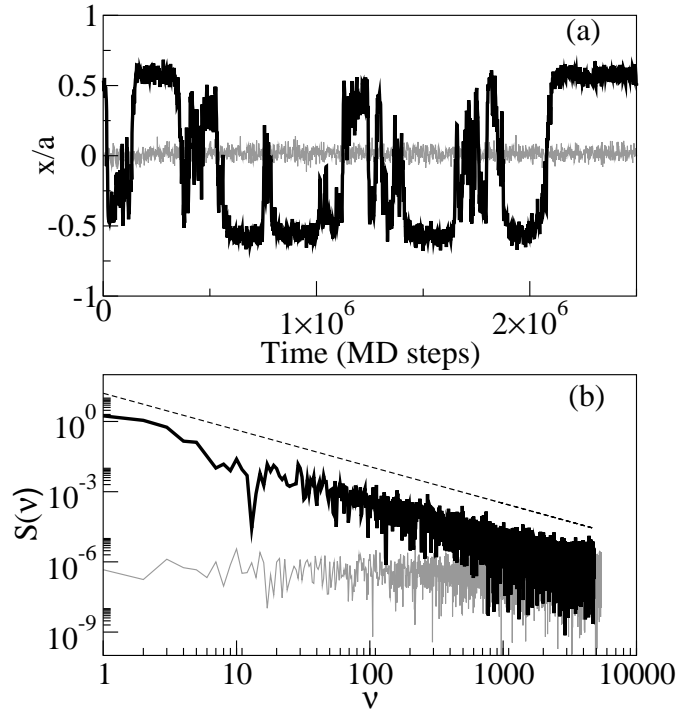}}
\caption{(a) Time series for the relative $x$-position of
a single vortex in the chain for 
$T/T_{m} = 0.15$ (light curve) and $T/T_{m} = 0.25$ 
(dark curve). 
For low $T$, the vortex is stationary, while for higher 
$T$, the vortex
jumps back and forth by about one chain lattice constant $a$. (b) The 
power spectra $S(\nu)$ for the full time series of the vortex position.
For $T/T_{m} = 0.15$ (lower light curve)
the noise spectra is of low power and white. 
For $T/T_{m} = 0.2$ (upper dark curve)
the noise power is much higher. The dashed line 
indicates a power law $1/f^{\alpha}$ with $\alpha = 1.6$.   
}
\label{fig2}
\end{figure}

\hspace{-13pt}
For low $T$ there is little motion of the vortices in the chain or the
bulk as seen in Figure 1(a).
In addition the incommensuration in the chain is stationary or pinned in a 
well defined location. 
For increasing temperatures there is a 
transition to a state in which the vortex chain exhibits 
a higher mobility than the 
bulk vortices, as seen in the vortex trajectory traces of Fig.~1(b)
at $T/T_{m} = 0.25$.

To elucidate the exact nature of the motion along the chain, in 
Fig.~2(a) we plot the position of a single vortex in the chain  
for the states in Fig.~1(a) and Fig.~1(b).
For the stationary chain the vortex remains immobile with only small 
thermal fluctuations in position of magnitude less than $a$.
For the mobile chain state the vortex
does {\it not} move continuously, but instead moves in 
discrete jumps of magnitude nearly $a$. 
These jumps do not occur throughout the chain simultaneously,
but run through the chain as a soliton like pulse with the
vortices jumping sequentially.
The discrete jumps occur due to the depinning of the 
incommensurations, which move randomly through the sample,
causing each vortex to jump one lattice constant
as the incommensuration passes over it.
The discreteness of the
jumps arises from the periodic potential
in which the chain vortices sit,
created by the stationary ordered bulk vortices.
Since the incommensurations
are much more weakly pinned than the bulk vortices, they depin at a 
much lower temperature. 

Figure~2(a) also shows that although there is motion of vortices 
along the chain, 
a single vortex does not diffuse but only moves
back and forth by a 
single lattice constant. 
This can be understood by considering a
single randomly moving incommensuration. 
When the incommensuration traveling in the positive $x$-direction
passes over a vortex, the vortex moves by $a$ in the positive
$x$-direction.  
The {\it same} incommensuration {\it cannot}, however, later cause the 
same vortex to move 
by an additional distance 
in the positive $x$-direction, 
but only in the negative $x$-direction when the incommensuration returns.
Thus a single randomly moving incommensuration produces no vortex diffusion 
along the chain but rather an oscillation.  
If a series of incommensurations all moving in the same direction pass
over a vortex, 
the vortex could move several lattice constants in one direction.
This is prevented because there is no net shear in the system.
The moving incommensuration sets up a strain field that prevents
the incommensuration from traveling large distances in the same direction.
As a result, the incommensuration does not diffuse freely through
the chain but tends to drift back and forth about its original starting
position.

Using a Voronoi construction, we find that the incommensuration in
the vortex chain coincides with a local defect composed of one 5-fold and
two 7-fold coordinated vortices.
It is the 1D motion of this defect that occurs in Fig.~1(b). 
The images in Ref.~\cite{Matsuda15} 
also indicate that the initial blurring in the chain 
occurs where there are 
localized topological defects
in the vortex lattice.  This agrees well with our finding that
it is the motion of defects that is occurring along the chains. 

In Fig.~2(b) we show the power spectra for the mobile 
and immobile chain. 
For the immobile chain, the noise spectrum is white,
while for the mobile chain, the low frequency noise
power is several orders of magnitude higher,
due to the 
correlated low frequency motions of the incommensuration, 
and there is now a $1/f^{\alpha}$
signature, with $\alpha = 1.6$. For $T > T_{m}$, when all the vortices
diffuse randomly, the noise spectra again 
becomes white.  

In simulations
where the chain is {\it commensurate}
with the bulk vortex lattice, such that $a = a^{\prime}$,
the chain and bulk vortex mobility are the {\it same}.  
Thus, the increased mobility of the vortices along the chain,
as shown in Fig.~1(b), is due strictly to an
incommensuration effect. 
When the vortex density along the chain is increased, lowering
$a^{\prime}$ and increasing the number of incommensurations, the
thermal depinning temperature of the chain decreases.  The
depinning and phase locking described below persist 
as the number of incommensurations increases.

A further probe of the relative pinning strength of the
incommensurations 
along the chain, compared to the 

\begin{figure}
\center{
\epsfxsize=3.5in
\epsfbox{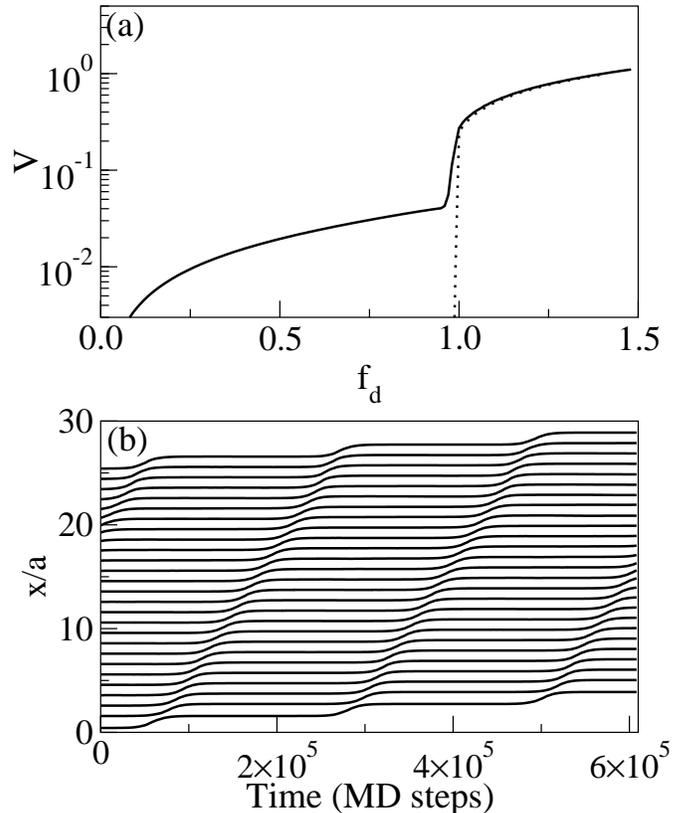}}
\caption{(a) Velocity $V$ vs applied dc drive $f_{d}$ for $f_{d}$ in the
$x$-direction at $T/T_{m} = 0.0$. 
In the first portion of the curve, for $f_{d}<1.0$, 
vortices along the chain have depinned while the bulk vortices 
remain pinned.   
(b) The consecutive vortex position along the chain vs time
for a fixed dc drive of $f_{d}/f_{p} = 0.15$. Each curve 
corresponds to the trajectory of a single vortex in the chain.  
} 
\label{fig3}
\end{figure}

\hspace{-13pt}
pinning of the bulk vortices,
is the application of a dc drive. 
We consider the case with weak random pinning where, 
in the 
absence of a chain, there is a single well defined 
elastic depinning transition. 
In Fig.~3(a) the velocity-force curve 
for the incommensurate case 
shows two depinning transitions;  the first is the depinning along
the chain, and the second at $f_{d}=1.0$ is the bulk vortex depinning.
A similar two stage depinning has been observed in 
artificial channels and periodic pinning arrays \cite{Kes8}.  
Imaging experiments in the vortex chain state also find that the 
chain depins first under an external driving force \cite{Matsuda15}. 
In Fig.~3(b) we plot the positions of the vortices in the chain
 as a function of 
time just above the depinning transition, showing that the motion is 
not continuous, but occurs in a
pulse or stick-slip motion.  Individual vortices remain stationary
most of the time, and then jump 
one lattice constant in sequence. The disturbance
or incommensuration moves much faster than the individual 
vortices. This is the same type of 
motion that occurs during the thermal depinning; however, in the driven case
the incommensuration moves in only one direction. A similar 
type of 
1D soliton motion has been 

\begin{figure}
\center{
\epsfxsize=3.5in
\epsfbox{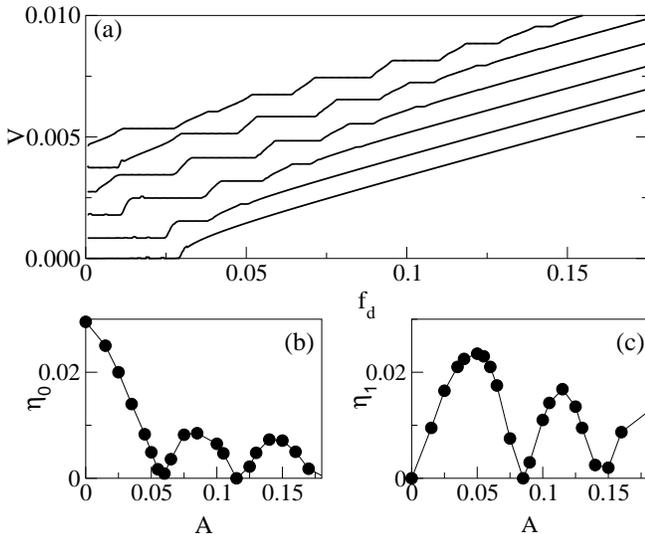}}
\caption{
(a) Velocity $V$ vs applied dc drive $f_{d}$ for 
$T/T_{m} = 0.0$ with an additional applied ac drive. 
The curves have been shifted up for presentation. Different 
ac amplitudes from bottom to top are $A =$ 0.0. 0.02, 0.04, 0.06, 0.1, and
0.12. (b) The width of the critical depinning force $\eta_{0}$ vs 
A. (c) The width of the first step $\eta_{1}$ vs A.  
}
\label{fig4}
\end{figure}

\hspace{-13pt}
seen for vortex motion in 2D periodic
pinning arrays
in simulations \cite{Reichhardt7} and experiments \cite{Surdeanu6} 
where the vortices effectively channel
along the pinning rows.   

We have also investigated the effects of a combined dc and ac drive on 
the moving chain state. Since the motion along the chain 
occurs over an effective periodic substrate, interference effects
should occur between the ac drive and the 
frequency induced by the dc drive. 
In this case, it is the incommensurations, rather than the
individual vortices in the chain,
that undergo phase locking.
In Fig.~4(a) we show a series of velocity-force curves for the moving chain
for different ac amplitudes and fixed
ac frequency. Here clear phase locking effects can be seen 
as steps in the velocity curves. The widths of the critical depinning
force and the higher order steps vary nonmonotonically with
ac amplitude. If the phase locking steps are of the Shapiro step type,
the $n$th step 
should vary as the Bessel function $J_{n}(A)$ \cite{Appl17}.
In Fig.~4(b) and Fig.~4(c), the widths of the depinning
force $\eta_{0}$ and the first step $\eta_{1}$ 
fit to $J_{0}$ and $J_{1}$, respectively,
showing that the phase locking is best described as a Shapiro step 
similar to those seen in sliding CDW's \cite{Gruner2}
as well as 2D vortex systems with 
periodic pinning \cite{VanLook5}.  The steps shown here occur {\it below}
the bulk depinning transition, for $f_{d}/f_{p}<1.0$, where only
the chain vortices are depinned.

In conclusion, we have investigated vortices confined in 1D embedded in 
an ordered 2D vortex lattice with weak pinning. 
The vortices in the chain 
are incommensurate with the periodic potential created by the bulk vortices. 
We confirm the conjecture in Ref.~\cite{Matsuda15} 
that the smearing of the vortex
chain state is due to an incommensuration effect, where for increasing
temperature the incommensurations thermally depin and
become mobile before the bulk vortices do.   
The thermal motion along the chain occurs through 
a soliton like pulse which 
moves randomly through the sample.
The vortices hop by one lattice constant
in the positive or negative direction as the pulse moves through. 
The onset of motion along the chain can also be observed as an increase
in the voltage noise with a $1/f^{\alpha}$ characteristic spectra. 
Under application of an external dc drive, we observe a two stage 
depinning where the
vortices in the chain depin first and form running solitons. 
When an ac drive is added to the dc drive, we observe phase 
locking of the moving 
incommensurations in the form of Shapiro steps 
in the velocity-force curves. 
Our predictions can be tested through
noise measurements in the vortex chain state under application of
external drives or temperature,
and also apply to vortex motion in artificial channels. 

{\it Note}: Just before submitting this work we
became aware of recent experiments that find phase locking of dc and ac 
driven vortices in artificial channels \cite{Nokubo18}.  

This work was supported by the US Department of
Energy under Contract No. W-7405-ENG-36. 
We thank P. Kes and A. Tonomura for useful discussions.


\vspace{-0.2in}
\begin{references}
\vspace{-0.5in}

\bibitem{Persson1}
B.N.J.~Persson, {\it Sliding Friction: Physical Principles and
Applications} (Springer, Heidelberg, 1998). 

\bibitem{Gruner2}
G.~Gr{\" u}ner, Rev.~Mod.~Phys.~{\bf 60}, 1129 (1988);
R.E.~Thorne, Phys.~Today {\bf 49}(5), 42 (1996). 

\bibitem{Giamarchi3}
L.~Balents, M.C.~Marchetti, and L.~Radzihovsky, Phys.~Rev.~B ~{\bf 57},
7705 (1998); 
P. LeDoussal and T.~Giamarchi, Phys.~Rev.~B {\bf 57}, 11356 (1998). 

\bibitem{Moshchalkov4}
M.~Baert {\it et al.}, Phys.~Rev.~Lett.~{\bf 74}, 3269 (1995);
K.~Harada {\it et al.}, Science {\bf 274}, 1167 (1996);
J.I.~Mart{\' \i}n {\it et al.}, Phys.~Rev.~Lett.~{\bf 79}, 1929 (1997).

\bibitem{VanLook5}
L.~Van Look {\it et al.}, Phys.~Rev.~B {\bf 60}, R6998 (1999); 
C.~Reichhardt {\it et al.}, Phys.~Rev.~B {\bf 61}, R11914 (2000). 

\bibitem{Surdeanu6}
R.~Surdeanu {\it et al.}, Europhys.~Lett.~{\bf 54}, 682 (2001). 

\bibitem{Reichhardt7}
C.~Reichhardt, C.J.~Olson, and F.~Nori, Phys.~Rev.~B {\bf 58}, 6534 (1998). 

\bibitem{Kes8}
A.~Pruymboom {\it et al.}, Phys.~Rev.~Lett.~{\bf 60}, 1430 (1988);
M.H.~Theunissen, E.~Van der Drift, and P.H.~Kes, {\it ibid.}~{\bf 77},
159 (1996); R.~Besseling, R.~Niggebrugge, and P.H.~Kes, {\it ibid.}~{\bf
82}, 3144 (1999); R.~Besseling {\it et al.}, cond-mat/0202485.   

\bibitem{Smith9}
S.~Anders {\it et al.}, Physica C {\bf 332}, 35 (2000);
Phys.~Rev.~B {\bf 62}, 15195 (2000).

\bibitem{Hog10}
M.J.~Hogg {\it et al.},  Appl.~Phys.~Lett.~{\bf 78}, 1433 (2001). 

\bibitem{Grain11}
A.~Gurevich {\it et al.}, Phys.~Rev.~Lett.~{\bf 88}, 097001 (2002).

\bibitem{Groth12}
C.A.~Duran {\it et al.}, Nature (London) {\bf 357}, 474 (1992);
Phys.~Rev.~Lett.~{\bf 74}, 3712 (1995);
U.~Welp {\it et al.}, {\it ibid.}~{\bf 74}, 3713 (1995);
J.~Groth {\it et al.}, {\it ibid.}~{\bf 77}, 3625 (1996). 

\bibitem{Bolle13}
C.A.~Bolle {\it et al.}, Phys.~Rev.~Lett.~{\bf 66}, 112 (1991). 

\bibitem{Bending14}
A.~Grigorenko {\it et al.}, Nature (London) {\bf 414}, 728 (2001).

\bibitem{Matsuda15}
T.~Matsuda {\it et al.}, Science {\bf 294}, 2136 (2001). 

\bibitem{Dodgson16}
M.J.W.~Dodgson, cond-mat/0201197. 

\bibitem{Appl17}
A.~Barone and G.~Paterno, {\it Physics and Applications of the 
Josephson Effect} (Wiley, New York 1982). 

\bibitem{Nokubo18}
N.~Kokubo {\it et al.}, cond-mat/0203352. 

\end{references}
\end{document}